\documentclass[traditabstract]{aa}
\usepackage{graphicx}
\usepackage{txfonts}
\usepackage{natbib} 
\usepackage{url}
\usepackage{setspace}
\bibpunct{(}{)}{;}{a}{}{,} 

\newcommand{\teff}{\ensuremath{T_{\rm{eff}}}}
\newcommand{\logg}{\ensuremath{\log g}}

\newcommand{\aador}{AA\,Dor}

\newcommand{\msol}{\ensuremath{M_{\odot}}}

\newcommand{\kirr}{\ensuremath{K_{\mathrm{irr}}}}
\newcommand{\ksd}{\ensuremath{K_{\mathrm{sdO}}}}
\newcommand{\msd}{\ensuremath{M_{\mathrm{sdO}}}}
\newcommand{\vsd}{\ensuremath{v_{\mathrm{sdO}}}}
\newcommand{\kdm}{\ensuremath{K_{\mathrm{2}}}}
\newcommand{\mdm}{\ensuremath{M_{\mathrm{2}}}}
\newcommand{\hbeta}{\ensuremath{H{\small \beta}}}
\newcommand{\hgamma}{\ensuremath{H{\small \gamma}}}

\newcommand{\hzeta}{\ensuremath{H{\small \zeta}}}
\newcommand{\htheta}{\ensuremath{H{\small \theta}}}
\newcommand{\phase}[1]{{\ensuremath{\varphi}\,=\,#1}}

\begin{document}

\title{Looking at the bright side -- The story of\\
       AA\,Dor\ \thanks{
       Based on observations collected at the European Southern Observatory,
       Chile. Program ID: 066.D-1800.} as revealed by its cool companion}

\author{
   M.~Vu\v{c}kovi\'{c}\inst{1} \and
   R.~H.~\O stensen\inst{2} \and
   P.~N\'{e}meth\inst{2,3} \and
   S.~Bloemen\inst{2,4} \and 
   P.~I.~P\'{a}pics\inst{2}
   }

\offprints{maja.vuckovic@uv.cl}

\institute{ 
  Instituto de F\'{i}sica y Astronom\'{i}a, Facultad de Ciencias, Universidad de Valpara\'{i}so, Gran Breta\~{n}a 1111, Playa Ancha, Valpara\'{i}so 2360102, Chile \\
   \email{maja.vuckovic@uv.cl}
\and Instituut voor Sterrenkunde, KU Leuven, Celestijnenlaan 200D, B-3001 Leuven, Belgium
\and Dr.~Karl-Remeis-Observatory \& ECAP, Astronomical Institute,
     F.-A.-U.\,Erlangen-N\"urnberg, 96049 Bamberg, Germany
\and Department of Astrophysics, IMAPP, Radboud University Nijmegen, PO Box 9010, 6500 GL, Nijmegen, The Netherlands      
}

\date{Received DD MM 2015 / Accepted }

\abstract{Irradiation effects in secondary stars of close binary systems are crucial for a reliable determination of system parameters and understanding the close binary evolution. They affect the stellar structure of the irradiated star and are reflected in the appearance of characteristic features in the spectroscopic and photometric data of these systems.
We aim to study the light originating from the irradiated side of the low mass component of close binary eclipsing system comprising a hot subdwarf primary and a low mass companion, in order to 
precisely interpret their high precision photometric and spectroscopic data, and accurately determine their system and surface parameters. 
We re-analyse the archival high-resolution time-resolved VLT/UVES spectra of \aador\ system where irradiation features have already been detected. After removing the predominant contribution of the hot
subdwarf primary, the residual spectra reveal more than 100 emission lines from the heated side of the secondary that show maximum intensity close to the phases around secondary eclipse. We analyse the residual spectrum in order to model the irradiation of the low mass secondary. 
We perform a detailed analysis of 22 narrow emission lines of the irradiated secondary, mainly of \ion{O}{ii}, with a few significant \ion{C}{ii} lines. Their phase profiles constrain the emission region of the heated side  to a radius $\geq$ 95 \% of the radius of the secondary, while the shape of their velocity profiles reveals two distinct asymmetry features one at the quadrature and the other at the secondary eclipse.
In addition, we identify weaker emission signatures originating from more than 70 lines including lines from \ion{He}{i}, \ion{N}{ii}, \ion{Si}{iii}, \ion{Ca}{ii} and \ion{Mg}{ii}. From the emission lines of the heated side of the secondary star we determine the radial velocity semi-amplitude of the center-of-light and correct it to the centre-of-mass of the secondary which in turn gives accurate masses of both components of the \aador\ system. The resulting masses $M_{1}$\,=\,0.46 $\pm$ 0.01\msol\ and $M_{2}$\,=\,0.079 $\pm$ 0.002\msol\ are in perfect accordance with those of a canonical hot subdwarf primary and a low mass just at the substellar limit for the companion. We also compute a first generation atmosphere model of the low mass secondary, which includes irradiation effects and matches the observed spectrum well. We find an indication of an extended atmosphere of the irradiated secondary star.   
}

\titlerunning{Looking at the bright side of the secondary of \aador}
\authorrunning{M.~Vu\v{c}kovi\'{c} et al.}

\maketitle


\section{Introduction}\label{sect:intro}

The booming research field of hot-subdwarf stars has been enriching our knowledge on very broad and diverse fields of astrophysics, all the way from stellar evolution and asteroseismology to planet findings (see \citealt{Ulisdbreview} for a review). However, these unique stellar laboratories offer further opportunities.  As a significant fraction of hot subdwarfs are found in binaries \citep{Maxted_2001,Napiwotzki2004SPY,Morales-Rueda2006ECSurvey, Copperwheat2011}, they present a superb stellar population to study binary evolution. In fact, the leading formation channels for these stars involve, in one way or another, binary evolution processes \citep{Han2002,Han2003}.

More than 100 binary hot subdwarfs have been found in short-period systems with periods ranging from a few hours to several days, with companions being either white-dwarf (WD) or a low mass M-dwarf stars (dM) \citep{Kawka2015}. Such short-period systems must  have formed through binary mass transfer followed by common-envelope ejection (CEE).  Due to their extremely thin ($<$ 0.02 \msol) and hence inert hydrogen envelope the subdwarfs evolve directly into WDs,  bypassing the asymptotic giant branch (AGB) and planetary nebula (PN) phases \citep{Heber1984, Saffer1994}. As the shortest period hot subdwarf + dM systems will evolve into cataclysmic variables (CVs) in less than a Hubble time, they are important for understanding the pre-CV evolution, with an observational advantage over the CV systems since their binary components are not hiding in the glare of an accretion disk. Hence, the short-period hot subdwarf binaries represent an important progenitor channel of CVs.  In fact, \citet{Schenker2005} postulate that the CVs below the period gap are the product of post-subdwarf binary evolution.   On the other hand, the hot subdwarfs with massive WD companions are potential progenitors of Type Ia supernovae (see \citealt{Iben1984, Webbink1984}, and for the most recent candidates see \citealt{Geier2013, Vennes2012}). Therefore, they are an important ingredient for understanding Type Ia supernovae.

Establishing the accurate masses of both components of post-common-envelope binaries (PCEBs)  is the only empirical way to constrain the efficiency parameters of the CEE phase.  Furthermore, detailed investigation of hot subdwarf binaries is  crucial in order to determine their masses for comparison with their theoretically proposed evolutionary channels as well as to test the binary-population synthesis models.
However, most of the hot subdwarf binary systems with known orbital periods are single-lined spectroscopic binaries, making it impossible to reliably determine the absolute masses of their components. Among the PCEBs there are 15 eclipsing binary systems found to date comprised of a hot subdwarf primary  (sdO or sdB star) and a low mass secondary, either a late dM or a brown dwarf (BD) (for the most recent discoveries see \citealt{Schaffenroth2013, Barlow2013, Schaffenroth2014eclipsing, Schaffenroth2015}). Such eclipsing systems are known as HW-Vir-type binaries after the prototype. In addition to eclipses HW-Vir systems are notorious for having a very strong `reflection effect'. This reflection effect, characteristic for all close binaries comprising a very hot primary and much cooler secondary, is due to the high contrast in the temperatures between the heated and unheated hemisphere of the secondary star.  As the secondaries in these systems are supposed to be orbiting synchronously, one side of the cool secondary star is constantly being irradiated by the strong ultraviolet (UV) flux of the hot-subdwarf primary. The light-curve solutions of HW-Vir-type systems imply that the temperature of the cool companion around the substellar point is increased up to 10\,000 -- 20\,000\,K \citep{Hilditch2003, Kiss2000}. However, the detection of spectral features from the irradiated secondary in HW-Vir-type  systems is hampered by the contamination from the much hotter subdwarf primary.

Recently, the eclipsing PCEBs gained even more attention based on the fact that nearly all systems with accurate eclipse-timing measurements having long enough data sets are found to show apparent orbital-period changes \citep{Kilkenny2014} that might indicate the presence of circumbinary planets \citep[see][for the most convincing cases]{Beuermann2010, Beuermann2012a, Beuermann2012b}. The claims of circumbinary planets in short-period sdB+dM/BD systems as well as various planets in close orbits around single subdwarf B stars \citep{Silvotti2007, Charpinet2011} show that planets not only may survive through the CE evolution, but also that they might play an important role in the formation of hot subdwarf stars \citep{Bear2011, Soker1998}.  Whether a planet in a PCEB system has survived the CEE or has been formed out of the ejected envelope (so-called second-generation planet, see \citealt{Zorotovic2013}) remains a question that current observations are just about starting to tackle. The nature of such substellar objects can be addressed through the analysis of their irradiated light, which in turn gives their masses.  

Interestingly, \citet{Kilkenny2014} showed  that \aador\ has stable orbital period to the level of $10^{-14}$\,d per orbit based on 33 years of primary eclipse timings, which was recently confirmed by analysing multi-year observations from the SuperWASP archive \citep{Lohr2014}. This makes \aador\ a unique eclipsing subdwarf PCEB system as all other HW-Vir-type systems measured with sufficient accuracy show semi-periodic variations in the eclipse timings.

In this paper we present a detailed study of the strong irradiation effect in \aador\,, a bona fide member of the HW-Vir-type systems, in which the irradiated light from the super-heated face of the secondary was first detected by \citet{AADormyIpaper}, and recently confirmed by \citet{Hoyer2015}. 


\section{The tale of \aador}\label{sect:aador}

Ever since \citet{Kilkenny1975MNRAS}  first reported the variations in the lightcurve of \aador\ (also known as  LB\,3459), a blue star in the foreground of the Large Magellanic Cloud never ceased to draw the attention.  Even though the \aador\ system is one of the most studied short-orbit subdwarf systems, the controversy over the nature of its components still remains.  The discussion on the evolutionary status of \aador\ started soon after the discovery with the works of \citet{Paczynski1980} and \citet{Kudritzki1982} who more-or-less agreed on the two most likely possibilities: that \aador\ is either a helium WD with a mass of about 0.3 \msol\ or a carbon-oxygen WD with a mass of 0.55 \msol\ with a degenerate red-dwarf companion with a mass of 0.04 \msol\ and 0.07 \msol\, respectively. However, these early speculations were disputed by \citet{Kilkenny1981MNRAS} and \citet{Hilditch1980MNRAS} who proposed a new evolutionary model with a mass ratio near 0.1 and a primary of 0.5 \msol\ with a 0.07 \msol\ secondary, concluding that the primary is a subdwarf O star.   

Determined to reveal the true nature of its companion, the first high-resolution (R\,=\,46\,890) time-resolved spectra of \aador\ were collected by \citet{Rauch2003}. The data set revealed the expected Rossiter--McLaughlin (RM) effect \citep{RMeffect} but it was not used to refine the parameters. 
 
The strong reflection effect seen in the lightcurve implies that the secondary component is strongly heated, however, many attempts to detect spectral features of the secondary star in \aador\ failed, e.g. \citet{Hilditch1996MNRAS,Rauch2003}. This inevitably makes it impossible to derive the absolute sizes of stars in this system, and so the speculation of the masses of the stars in \aador\ continued. In a single lined binary system only the mass function can be determined, and one is forced to make sound assumptions on the mass of one star in order to determine the absolute sizes of both stars. 
As the derived surface gravity and effective temperature of the primary fall into the extended horizontal branch (EHB) region, where the hot subdwarf stars reside, the sound assumption would be that the primary mass is the canonical mass of a hot subdwarf star. This was done by \citet{Hilditch2003}, who imposed the 0.5\msol\ for the hot subdwarf primary star and from there derived a mass of the secondary of 0.09 \msol, which agrees with an M-dwarf with a surface temperature of about 2\,000\,K.  However, there is another group of stars that crosses the EHB region of the \teff\ -- \logg\ diagram on their way to the WD graveyard; the post-red giant branch (RGB) stars. These stars have been formed by extreme mass loss on the RGB before they could ignite helium. Due to this extreme mass loss they cannot reach the density required for the core helium flash, and evolve directly into low-mass WDs \citep{Driebe1998}. By comparing the location of \aador\, as derived from their high-resolution optical and UV spectra, with the evolutionary tracks for post-RGB stars \citet{Rauch2003} assumed a mass of 0.33\,\msol\ for the \aador\ primary, which consequently gave a correspondingly lower mass for the secondary component of 0.066\,\msol -- below the substellar limit \citep{Baraffe2015}.
 
Since \aador\ is one of the few systems where accurate mass determination is possible, these numbers are actively used to constrain evolutionary models \citep{Zorotovic2010}.  With the aim to search for signatures of the irradiated secondary star, which would in turn provide us with the accurate masses of the \aador\ system,  \citet{AADormyIpaper} re-analysed the VLT/UVES data set of  \cite{Rauch2003} discovering more than 20 emission lines originating from the heated side of the secondary. Based on a lower limit of the radial velocity of the irradiated object \citet{AADormyIpaper} made a first estimate of the masses of \aador\ components which were consistent with a regular EHB primary and a low-mass M-dwarf secondary, contrary to the results of \cite{Rauch2003}.

Refining the non-LTE models and including the proper treatment of the irradiation \citet{KleppyRauch2011} re-analysed their VLT/UVES data set and found a full agreement with the photometric model of \citet{Hilditch2003}. This solved, once and for all, the long-lasting  \logg\ problem. The mass they found for the primary, by comparing their spectroscopic solution to the evolutionary tracks of \citet{Dorman1993},  rules out the post-RGB scenario and is in full accordance with the one found by \citet{AADormyIpaper}. They derived the secondary mass of 0.079$\pm$\,0.007\msol\, which once again confirmed that the secondary is at the hydrogen burning mass limit, i.e. either a brown dwarf or a late M-dwarf star. 

While we were finalising this paper one more article was published on the same star. Armed with fresh XSHOOTER phase-dependent spectroscopy \citet{Hoyer2015} combine their FUSE \citep{Fleig2008} and UVES \citep{Rauch2003} data to yet again search for the signatures of the low mass secondary in the \aador\ system.  They identified 57 emission lines of the secondary star in the UVES spectra, and 53 spectral lines in the XSHOOTER spectra. Out of these they used 22 lines (14 from UVES data, 7 from XSHOOTER UVB arm and 1 from XSHOOTER VIS arm) to measure the orbital center-of-light velocity of the secondary star. Assuming that the centre of gravity of the measured irradiated light is dominated by a region lying towards the primary star they estimated the radial velocity semi-amplitude of the secondary to be $232.9^{16.6}_{-6.5}$\,km/s and correspondingly derive the masses of the both stars in \aador\, $M_{\mathrm{1}}$\,=\,$0.475^{0.0975}_{-0.0357}$\,\msol\, and $M_{\mathrm{2}}$\,=\,$0.0811^{0.0184}_{-0.102}$\,\msol\, which are in full agreement with their previous result \citep{KleppyRauch2011} as well as with the results of \citet{AADormyIpaper}.

  
\begin{figure}
\centering
\includegraphics[width=\hsize]{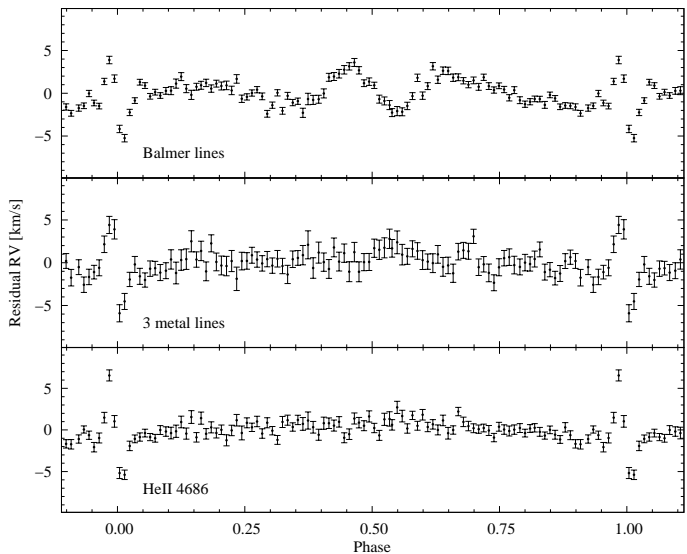}
\caption{The orbit subtracted RV residuals of the primary with their corresponding errors. Top panel is for the 3 Balmer lines H$_{\delta}$, H$_{\gamma}$ and H$_{\beta}$ lines. Middle panel is for the metal lines O III at 3759.87\AA{}, Si IV at 4088.86\AA{} and 4116 \AA{} and Mg II at 4481.23\AA{} and the bottom panel for HeII at 4685.79\AA{}.}
\label{Fig_RMresiduals}
\end{figure}


\section{Wiggles and wriggles in the velocities}

While examining the orbit-subtracted radial-velocity (RV) residuals in the UVES spectra of NY\,Vir  \citet{Vuckovic2007AA} noticed an apparent distortion at orbital phases around the secondary eclipse. As this feature occurs right where the contribution from the irradiated side of the secondary star is at its maximum it could therefore be an artifact of the reflected light or a fingerprint of the heated face of the secondary.
Due to complexities caused by the pulsating primary of the NY\,Vir system and the low S/N of this particular data set, \citet{Vuckovic2007AA} were not able to analyse this effect any further.
Motivated by this intriguing finding we then decided to search for spectra of a similar system in which such a feature could emerge. The high S/N time-resolved UVES spectra of \aador\ obtained by \citet{Rauch2003} were ideal for this purpose. We retrieved this data set from the ESO archive and reduced them using the standard ESO CPL\footnote{CPL stands for Common Pipeline Library} routines.
While the individual spectra are of good quality with a typical S/N of the order of 50 per wavelength bin, the known problem with the imperfect echelle order merging\footnote{For a proper merging of the adjacent echelle orders, the blaze function has to be corrected. Unfortunately, in the standard UVES pipeline reduction, the blaze function is estimated from a flat-field exposure, which is obviously not able to properly correct the blaze function.} is causing ripples in the wings of Balmer lines and the continuum level. As this is potentially fatal for our analysis and there was no way around it, we devoted special attention when normalising the spectra. 

We performed a half-automatised, two step normalisation of each spectrum, following the procedure of \citet{Papic2012}. All spectra were brought to the same flux level by dividing with their median intensity and we defined a master function (using the first spectrum) with cubic splines which was fitted to the continuum at fixed wavelength points known to be free of spectral lines. 
We then used this function to clean all spectra from large-scale artificial features, e.g.~the wiggles in the continuum caused by imperfect flat-fielding. The careful selection of these nodal points enabled us to construct a function which does not distort the shape (and especially the wings) of the Balmer lines.
To correct for small-scale effects and any remaining global trend of an instrumental origin, we carried out a second normalisation using many more nodal points, which were connected by linears at this phase. The wavelengths of these points were the same for all spectra, but the normalising function was constructed for each of them based on the local flux levels, instead of applying a master function again. Using linears at this step enabled us to leave the shape of wide features -- like the Balmer-wings -- untouched (by placing only one nodal point to the beginning and the end of such regions), but still make small-scale corrections (by using a denser nodal point distribution where it was needed). The described process is much faster, less subjective, and provides better consistency than a one-by-one manual normalisation.

\begin{table}[b]\small\rm
\caption[]{Fits to the radial velocity measurements.}
\label{tbl:rvs}
\centering
\begin{tabular}{lcc} \hline\hline \noalign{\smallskip}
Spectrum   & \ksd\  & $\gamma$ \\
           & [km/s]  & [km/s] \\
\noalign{\smallskip} \hline \noalign{\smallskip}
Balmer lines & --41.81\,$\pm$\,0.22 &$-$0.88\,$\pm$\,0.16 \\
\ion{He}{ii}\,4686 & --39.64\,$\pm$\,0.18 & +1.00\,$\pm$\,0.15 \\
3 metal lines& --39.63\,$\pm$\,0.21 & +0.97\,$\pm$\,0.13 \\
\noalign{\smallskip} \hline \noalign{\smallskip}
Adopted      & --39.63\,$\pm$\,0.21  & +0.97\,$\pm$\,0.13 \\
\noalign{\smallskip} \hline
\end{tabular}
\end{table}

\begin{figure*}
\centering
\includegraphics[angle=0, width=15cm]{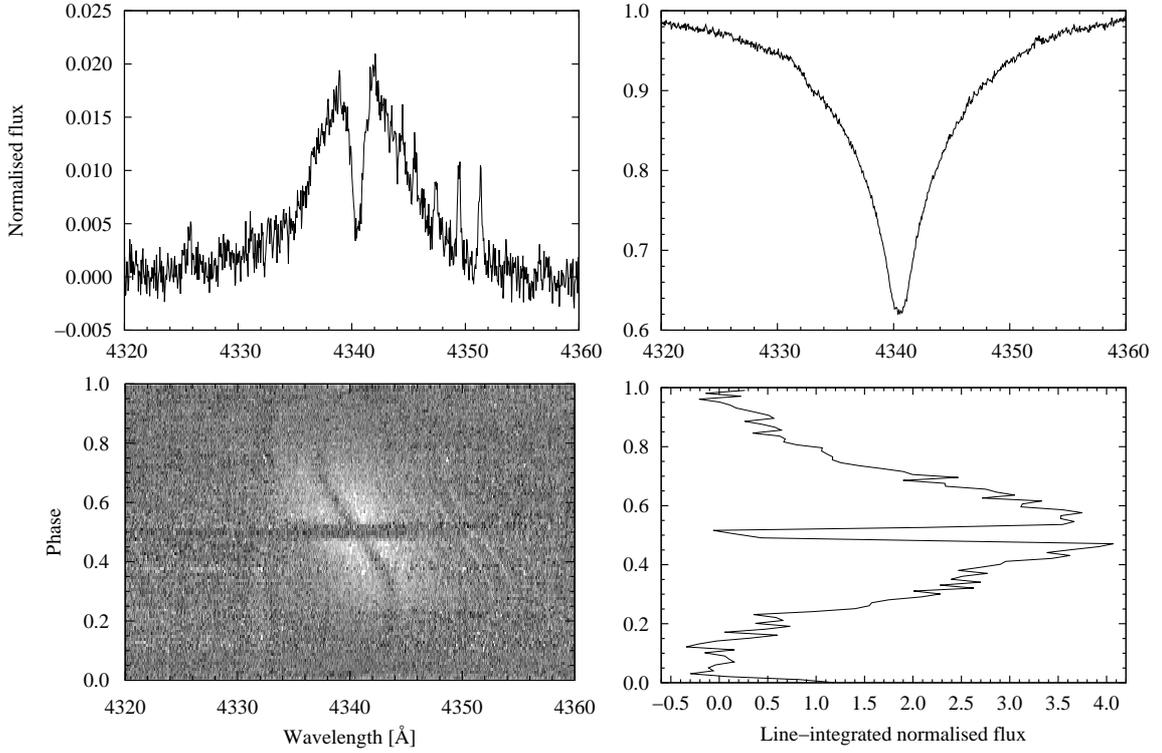}
\caption{The grayscale plot (lower-left panel) shows the phase folded spectra in the
         region around \hgamma, after shifting to the rest frame of the primary and
         subtracting the mean spectrum (upper-right panel) computed from the least
         contaminated phases. After shifting to the rest frame of the secondary and
         summing all phases, the emission profile of the secondary emerges
         (upper-left panel).
         Summing each phase-binned spectrum in a window 20\,\AA\ wide,
         and centered on \hgamma\ in the rest frame of the secondary, produces the
         phase profile shown in the lower-right panel.
        }
\label{fig:hgamma}
\end{figure*}

The sdO primary of the \aador\ system has a richer spectrum compared with the spectrum of the sdB primary in the NY\,Vir system. In addition to the broad Balmer absorption lines the sdO star displays a strong \ion{He}{ii} line as well as several clear metal lines which we used to measure the RV of the primary at different orbital phases.
We calculated the RVs by fitting a single Gaussian to each of the He and metal lines, while multi-Gaussians were fitted to the broad Balmer lines, following the procedure explained in \cite{Morales-Rueda2003}. The depth of the \ion{He}{ii} line and the Balmer lines were kept as free parameters in our fits, while the depth of the metal lines were fixed at their average value.
In this way we obtained the RVs of the Balmer lines from H$_{\beta}$ to H$_{\eta}$, of the \ion{He}{ii} line at 4685.79\AA\ and of several metal lines, like \ion{O}{iii} at 3759.87\AA, \ion{Si}{iv} at 4088.86\AA\ and \ion{Mg}{ii} at 4481.23\AA. 
To each of these sets of RV measurements we fitted a sine function locked in phase to the well-known ephemeris of the system \citep{Kilkenny2014}, leaving only the RV amplitude, \ksd, and the systemic velocity, $\gamma$, as free parameters.
The results from these sine fits to the three sets of RV measurements are given in Table\,\ref{tbl:rvs}.
The residual RVs that remain after subtracting the respective sine functions are plotted in Fig.\,\ref{Fig_RMresiduals} as a function of orbital phase, $\varphi$, where the zero phase is defined to correspond to the primary eclipse.
In addition to the well known Rossiter-McLaughlin (RM) effect, first reported in these spectra by \citet{Rauch2003}, we clearly see a broad distortion feature appearing in the Balmer lines just around the secondary eclipse (\phase{0.5}), whereas it is not present in any of the metal lines nor in the \ion{He}{ii} line. The fact that it appears only in the Balmer lines and not in the \ion{He}{ii} nor in the metal lines of the sdO star is a clear indication that this distortion is caused by the phase-dependent contamination of the Balmer lines by the corresponding lines present in the light from the irradiated hemisphere of the secondary star. 

\section{Signatures from the hot side}\label{sect:irradiation} 

After carefully reprocessing and normalising the UVES spectra as described
in the previous section, we proceeded by computing a mean spectrum of the primary taken
from phases close to the primary eclipse, but avoiding those affected
by the eclipse. The mean spectrum was constructed from nine individual spectra occurring
at phases, \phase{0.03\,--\,0.08}, just before and after
the primary eclipse when the cold side of the secondary is facing us.
The spectra were shifted by
\begin{equation}\label{eq:vsd}
\vsd = \ksd \sin(2\pi\varphi)
\end{equation}
before averaging.
The upper-right panel of Fig.\,\ref{fig:hgamma} shows the mean spectrum around the
\hgamma\ line.
Each of the 105 individual spectra were then cleaned from the
contribution of the sdO primary by subtracting the mean spectrum shifted according to Equation~\ref{eq:vsd}.
The residual spectra around the \hgamma\ line are shown in
the lower-left panel of Fig.\,\ref{fig:hgamma}. The trailed residuals clearly
show a very fast moving profile that changes velocity in antiphase with
the primary and has a maximum intensity just before and after \phase{0.5}
(secondary eclipse).
Superimposed on the red wing of the  \hgamma\  profile there are a number of
sharp emission features that all can be identified with \ion{O}{ii} lines.
Shifting by a velocity of \kirr\,$\approx$\,222\,km/s (see Section~\ref{sec:Kirr}) and
summing the contribution between \phase{0.23 - 0.47} produces the
spectrum shown in the upper-left panel of Fig.\,\ref{fig:hgamma}.
The lower-right panel of Fig.\,\ref{fig:hgamma} shows the phase profile
of this line, which is the total flux per phase bin in a 10\,\AA\ window
centered on the emission profile of each spectrum.

\begin{figure*}[ht!]
\centering
\includegraphics[angle=0,width=\hsize]{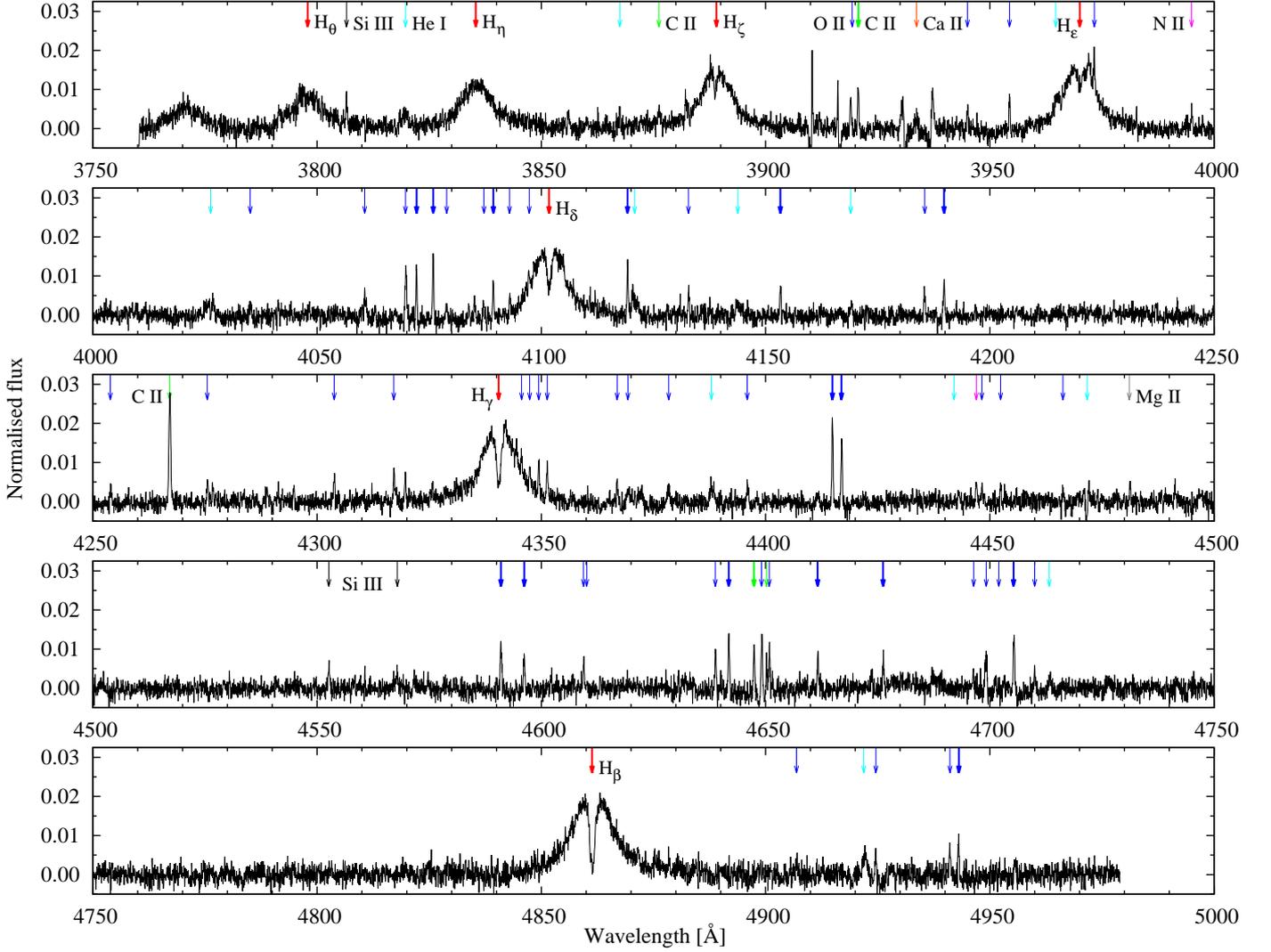}
\caption{Emission line spectrum of the secondary. This figure was generated
         in the same way as for
         the top left panel of Fig.\,\ref{fig:hgamma}, but for the
         full spectroscopic range. Lines used in our analysis
         are marked with thick arrows, and weaker lines with possible
         identifications listed in Table~\ref{tbl:emlines} are marked
         with thin arrows. (In the on-line version the color coding on
         the arrows identifies their species; Balmer lines: red,
         \ion{O}{ii}: blue, \ion{C}{ii} and \ion{C}{iii}: green,
         \ion{He}{i}: cyan, \ion{N}{ii}: purple, \ion{Si}{iii}: black,
         \ion{Mg}{ii}: gray.)
        }
\label{fig:dMspectrum}
\end{figure*}

The full emission-line spectrum of the secondary component is given in
Fig.\,\ref{fig:dMspectrum}. The arrows indicate the lines we have identified,
while thick arrows indicate those that were used in our radial-velocity analysis. 
All the Balmer lines from \hbeta\ to \htheta\ are clearly visible and
marked with labeled arrows in the figure. \hbeta\ to \hzeta\  appear to have
profiles with reversed cores.
As \citet{Barman2004} explained, these reversed core profiles are produced
by departures from local thermodynamic equilibrium (LTE) in 
layers above where the underlying emission line forms. 

In Table~\ref{tbl:emlines} we provide a list of all the metal lines that
originate from the irradiated face of the secondary component that
we could identify in the emission line spectrum. We list 
peak intensity and equivalent width for each of the 22 metal lines
that we have retained for use in the further analysis. 

\begin{figure*}
\centering
\includegraphics[angle=-90, width=16cm]{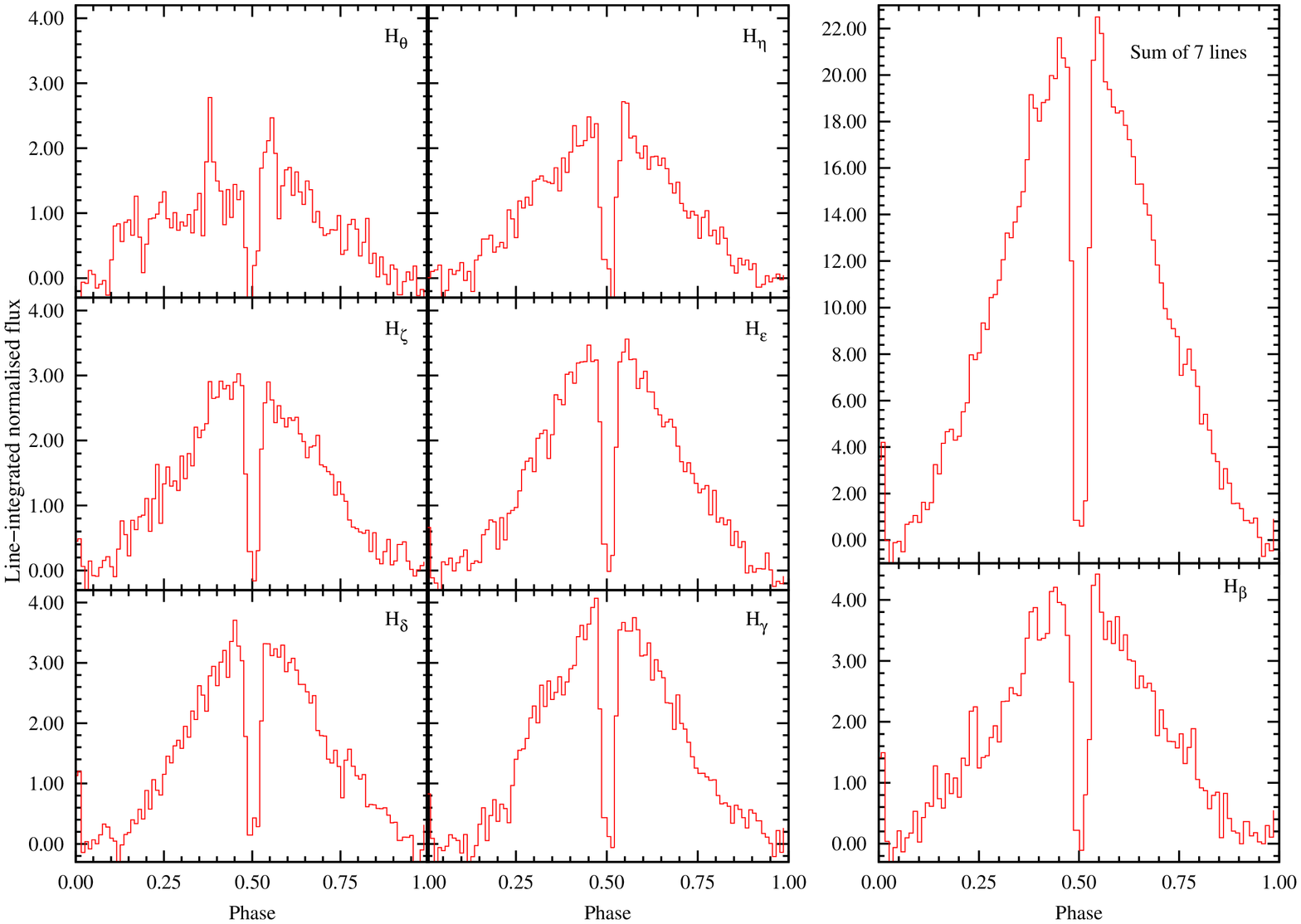}
\caption{Phase profiles for all the Balmer lines produced in the same way as for
         \hgamma\ in the lower-right panel of Fig.\,\ref{fig:hgamma}, and
         the sum of all 7 Balmer lines (upper-right panel).
        }
\label{fig:bprof}
\end{figure*}

\subsection{Emission line profiles}\label{sec:Kirr}

The phase profile of the emission lines is essentially the light curve
of the irradiated star. As can be seen from the lower-right panel of
Fig.~\ref{fig:hgamma} it shows a total eclipse at \phase{0.5} and
has a minimum that coincides with the primary eclipse.

Phase profiles for the strongest Balmer lines (computed as for \hgamma\ in
the lower-right panel of Fig.\,\ref{fig:hgamma}) are shown in Fig.\,\ref{fig:bprof},
as well as for the sum of all seven lines in the upper-right panel.
The shape of the phase profiles provides information on the size of the
emission region on the heated hemisphere of the secondary. There are no
significant differences between the different Balmer lines, so we use
the sum of the seven lines to analyse the phase dependence of the light
from the irradiated side of the companion.
It can be seen that the hot spot must cover almost the entire
hemisphere of the secondary since it only reaches
zero intensity immediately before and after the primary eclipse (\phase{0}).
In order to quantify this effect and determine the relationship between
the radial velocity measured from the irradiated hemisphere, the center-of-light velocity, \kirr\, and the
velocity of the center of mass of the secondary, \kdm\,
we must compute some simple geometrical models. In Section~\ref{sect:K2modeling}
below we compare the integrated Balmer-line profile with these models.
While the Balmer-line profiles are very broad in velocity space, the metal
line profiles are sharp enough to resolve the rotational profile of
the irradiated star in velocity. We summed up the metal line profiles as we
did for the Balmer lines. They show the same profiles in phase space, but with a much less signal than the Balmer-line profiles. In
velocity space, however, they reveal some even more interesting features.

\begin{figure*}
\centering
\includegraphics[width=\hsize]{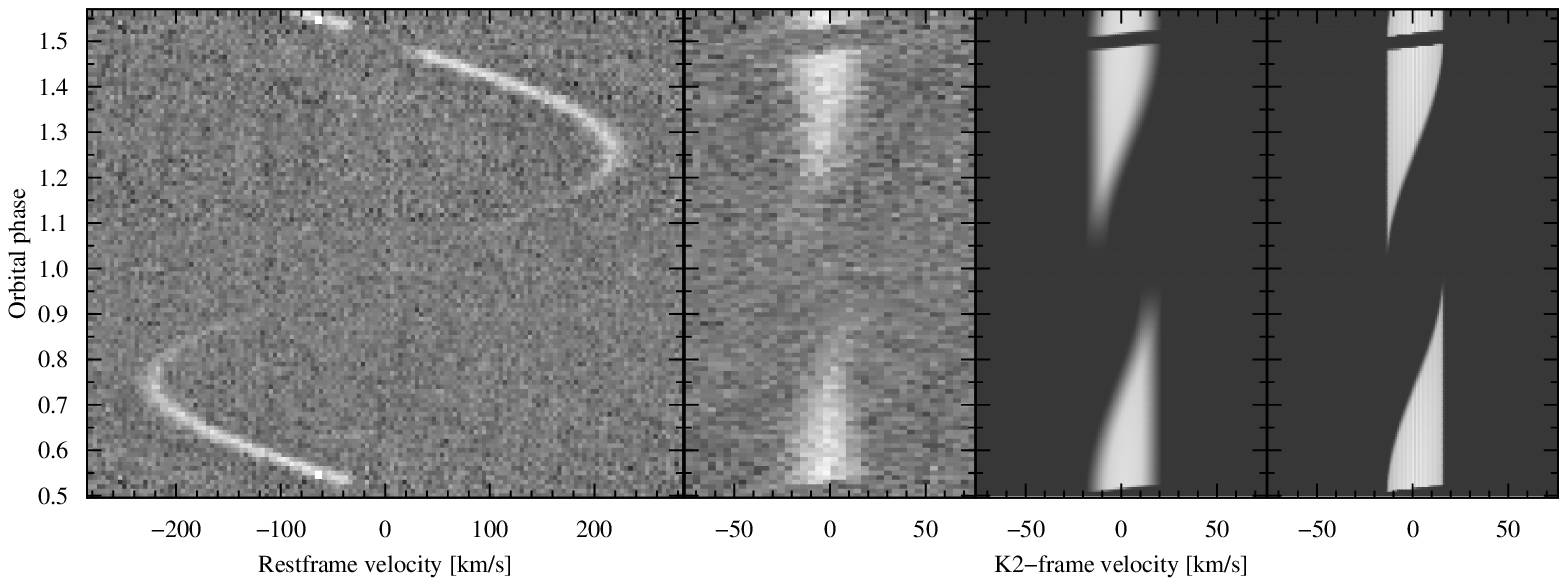}
\caption{Grayscale plot of the sum of all the metal lines in the rest
frame of the system (left) and shifted to the rest frame of the secondary
(middle). The last two panels show a geometrical model line profile for
comparison, with the first having been subjected to a Gaussian broadening.}
\label{fig:vprof}
\end{figure*}

The sharpness of the metal-line profiles makes them well suited for determining
the centre-of-light velocity, \kirr.
To enhance the signal as much as possible, we used the theoretical central
wavelengths of the 22 strongest lines (see Table~\ref{tbl:emlines}) and
added them up in velocity space. The resulting velocity/phase
image is shown in the left panel of Fig.\,\ref{fig:vprof}.
In order to find the optimum value for \kirr, we shifted the spectra by
a range of velocities in steps of 0.5 km/s,
and computed the integrated intensity for the 22 lines in a 0.1\,\AA\ window.
The maximum intensity is then found for 222.5\,km/s.
Fitting a parabola to the central 10\,km/s around the maximum of the
peak intensity profile gives the value we adopt for the center-of-light velocity,
\kirr\,=\,222.76\,$\pm$\,0.07\,km/s.

The second to the left panel of Fig.\,\ref{fig:vprof} shows the same spectra as in the
left panel, after shifting these according to the centre-of-light
velocity, \kirr.  Scrutinising this image we notice some
deviations from a linear symmetry. The profile appears to
be skewed towards a higher velocity (away from the observer) after
the final quadrature point \phase{0.75} and towards a lower
velocity (towards the observer) before the first quadrature.
This effect can be readily understood from geometry.
When the hot spot rotates into view after primary eclipse it first
appears on the side that rotates towards us, and so has a negative
radial velocity. We only see a complete rotationally broadened
line profile when the spot is completely in view, just before
and after secondary eclipse. As the hot spot rotates back onto the
far side, the dark side produces a self-eclipse effect that
reveals itself as a progressive dimming
of the approaching side of the line profile.
The last two right hand panels of Fig.\,\ref{fig:vprof} show a model line profile
based on the stellar radii derived by \citet{Hilditch2003} and our
radial velocity measurements, see Table~\ref{tbl:rvs}. The third panel shows a model line broadened by a 10 km/s Gaussian kernel
while the unbroadened one is presented in the fourth panel.
It is clear that the model reproduces all the distortion features seen in the observed integrated line profile.
The figure also visualises why the measured \kirr\ must be less than
the true centre-of-mass velocity \kdm, since the self-eclipse effect
always darkens the line-profile on the side that has the highest velocity
in the restframe picture.

Another asymmetry feature that emerges in the
emission-line profile is the direction of the eclipse
of the hot spot in velocity space during secondary eclipse
(best seen at the very top of the right panel in Fig.\,\ref{fig:vprof}).
At the beginning of secondary eclipse, at \phase{0.48}, we can see
that the negative part of the emission profile is obscured,
as the primary must cover the side of the secondary that is rotating
towards us (negative direction) before it reaches the half that
rotates away from the line of sight. At \phase{0.52} the opposite
occurs as the side that is rotating towards us emerges before
the receding side. This is of course the equivalent of the RM
effect for the hot spot on the irradiated secondary. 
 
\subsection{Determining $K_\mathrm{dM}$}\label{sect:K2modeling}

\begin{figure}
\centering
\includegraphics[angle=-90, width=\hsize]{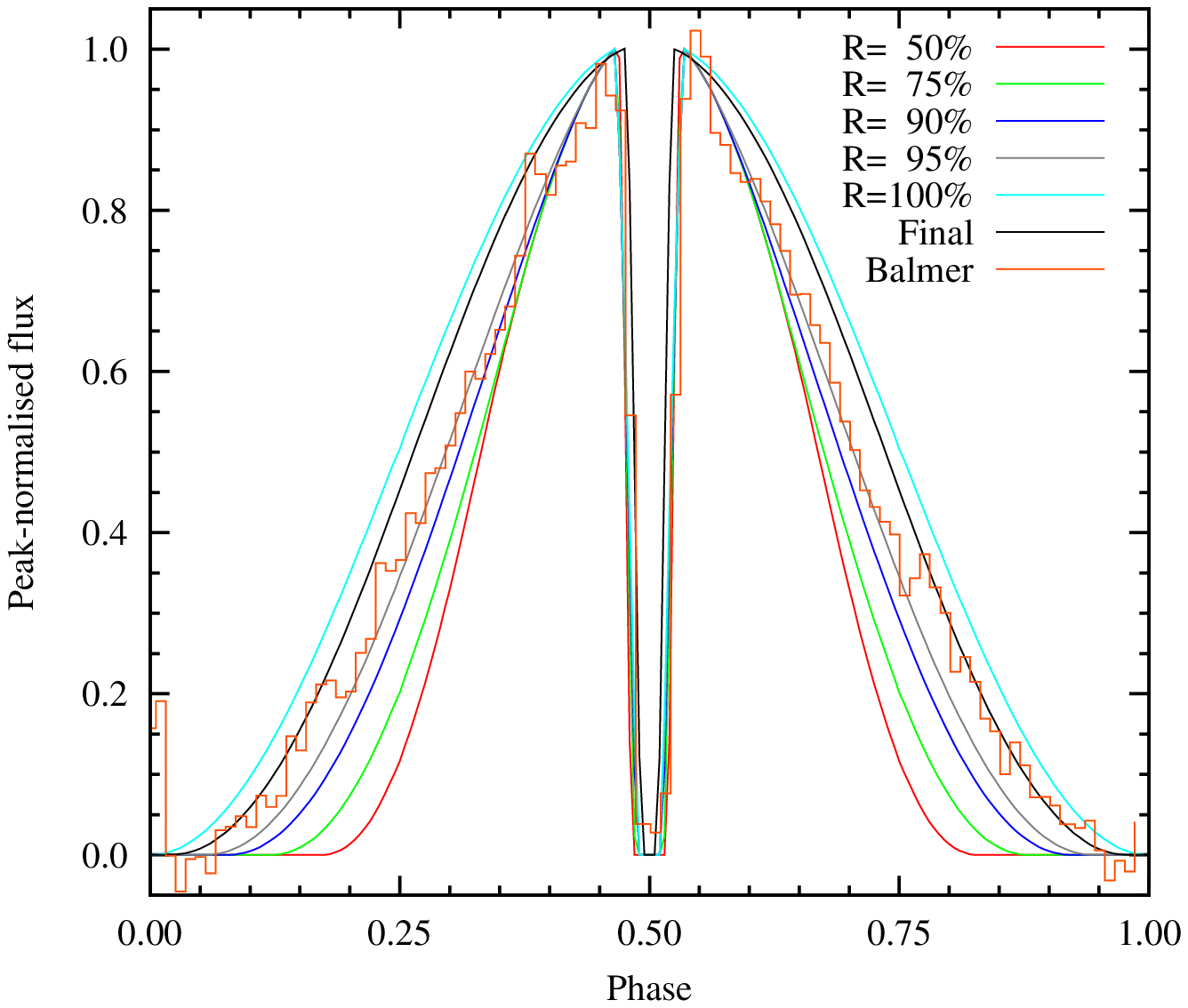}
\caption{Normalised phase profiles from models of emission
         from hot spots of various radii (curves). The `final' model
         includes edge effects and limb darkening.
         Also shown are the observed phase profiles from the sum of
         Balmer and metal lines.
         It is clear from comparing the model and observed profiles
         that the size of the hot spot must be between 95 and 100\%.
        }
\label{fig:Kcorr}
\end{figure}

In order to simulate the spectroscopic features of an irradiated star
that are geometrical in origin, we computed some simple numerical models.
In addition to the line-profile models shown in the previous section
we computed phase profiles for several different configurations.
The models consist of two bodies with the appropriate dimensions,
in a circular eclipsing orbit using the parameters for the
\aador\ system as determined by \citet{Hilditch2003}.
The surface of one body is taken to be dark (the primary), and the other
one is dark except for a circular patch with variable radius centered on
the axis between the two bodies.
In the simplest models we use a straight cutoff radius at a fraction of the
radius of the star, while in the `final' model we tapered the intensity of
the light towards the edge of the star so that the brightness of each
point was proportional to the area of the irradiating star that is visible
above the horizon for that point at a given orbital phase.
The phase profiles produced by the simulation were normalised
so that they peak at unity just before the start of the eclipse.
The resulting phase profiles for a hot-spot radius of 50, 75, 90, 95 and 100\%
are shown in Fig.\,\ref{fig:Kcorr}, together with the `final' model and
the integrated Balmer-line phase profile (from the upper-right panel of
Fig.\,\ref{fig:bprof}).
Clearly, a hot spot that covers 75\% or less of the irradiated hemisphere
disappears behind the rim of the companion around \phase($\pm$0.35),
much sooner than in the observed profiles. Only profiles of a hot spot covering
between 95 and 100\% of the irradiated hemisphere are compatible with
the observations. The `final' model seems to fit perfectly at phases
close to primary eclipse, and not so well closer to secondary eclipse.
But note that since there is some arbitrariness regarding the normalisation
especially of the peak level at \phase{$\sim$0.5}, and also the continuum level
close to \phase{0}, the apparent precision of the fit may be deceiving.

From the irradiated-light models we can also compute the difference between
the the center-of-light velocity and the center-of-mass velocity.
For a simple hot spot with a radius of 0.975, we find
\kirr\,=\,0.956\,\kdm. For the `final' model we can use the same procedure
that we used to derive \kirr\ from the observational data. We take the
line-profile model shown in the right-hand panel of Fig.\,\ref{fig:vprof}
and reintroduce a velocity on the order of \kirr, after which we locate
the velocity shift that produces the maximum intensity of a narrow
window across the line. We find that the measured velocity is 0.957
times that introduced to the geometrical model. However, if we use
the model profile subjected to Gaussian broadening, we get a slightly smaller
reduction factor; 0.963. We take the latter value as the most likely
value for the correction factor, \kirr/\kdm. The error on this number
is not easy to quantify. From the differences between a sharp and broadened 
model we estimating it to be on the order of 0.3\% and conclude
that the center-of-mass velocity for the secondary is
\begin{equation}
\kdm = \frac{\kirr}{0.963 \pm 0.003} = 231.3 \pm 0.7\textrm{km/s}.
\end{equation}
This results in a mass ratio
\begin{equation}
q = \frac{\kdm}{\ksd} = 0.171 \pm 0.001.
\end{equation}
The masses for the two components then follow,
\begin{eqnarray*}
\msd & = & 1.036\cdot10^{-7}  \left( \frac{\kdm}{\sin i} \right)^3 ( q + 1 )^2 P  =  0.46 \pm 0.01 \msol \\
\mdm   & = & q\,\msd  =  0.079 \pm 0.002 \msol
\end{eqnarray*}
implying a canonical mass for the hot subdwarf and a mass just at the
substellar limit for the companion \citep{Baraffe2015}. 

\begin{figure*}[ht!]
\centering
\includegraphics[angle=0,width=\hsize]{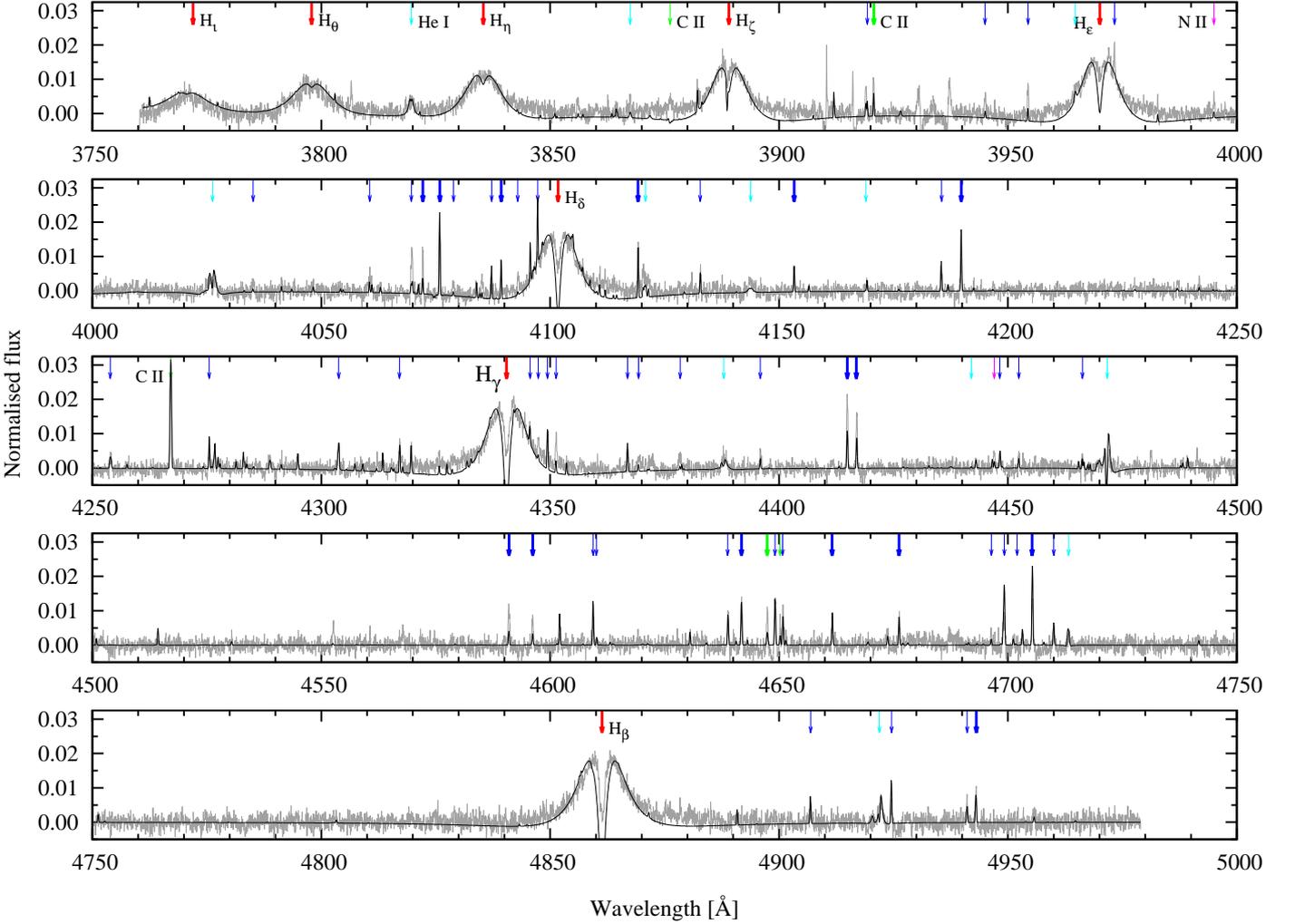}
\caption{Synthetic emission line spectrum of the dM secondary. Lines marked coincide with the ones shown in Fig.\,\ref{fig:dMspectrum}.
         (In the on-line version the colour coding on the arrows identifies their species; Balmer lines: red,
         \ion{O}{ii}: blue, \ion{C}{ii} and \ion{C}{iii}: green,
         \ion{He}{i}: cyan, \ion{N}{ii}: purple)
        }
\label{fig:TlustydMspectrum}
\end{figure*}

\section{Modelling the spectrum of the irradiated secondary}\label{sect:irradiated dM spectrum}

Irradiation effects are strong in close, HW-Vir-like subdwarf binaries with
cool companions. 
In non-eclipsing binaries the reflection effect usually completely dominates 
the light curve. 
In the AA Dor system the primary is a hot sdO subdwarf, and hence the
companion is subjected to a more extreme UV irradiation. 
Light curve modelling codes like the Wilson-Devinney variants 
(e.g. \citet{WDcode1971} or {\sc PHOEBE} \citep{Degroote2013}  
assume that the flux variation is due to reflection off the surface 
of the secondary or they include some form of energy redistribution 
on the companion. 
While this approach can be conveniently parametrised with 
only a few 
parameters and gives a very good quantitative description of the 
light curve, it does not provide the physical conditions in the 
atmosphere of the companion. 
To fully understand the evolution of close binaries
we must investigate the reflection effect. 
By understanding the radiative interactions 
between the members of close binaries we can derive
accurate surface 
parameters and better interpret their high precision
photometric data, such as the {\it Kepler} eclipsing binary 
light curves \citep{Prsa2011}. 
This methodology will eventually lead to the characterisation of 
non-transiting exoplanets based on spectral models and measured 
reflection light curves.

To model the effects of irradiation and reproduce both the spectral variations 
and the reflection effect in the light 
curve of AA\,Dor we followed the decription on the geometry
of eclipsing binaries developed by \citet{Gunter2011} 
and \citet{Wawrzyn2009}. 
This model provides the visible fractions of equally irradiated zones
on the companion as a function of orbital phase. 
For compatibility with \citet{Gunter2011} we adopted the same spherical 
coordinate system with the origin at the center of the secondary star 
and the z-axis intersecting the center of the primary. 
The z-axis intersects the surface of the companion at the substellar point, 
where irradiation is the strongest. 
Constant latitudes $\theta$ on the secondary (measured from the z-axis) 
receive the same amount of incident flux in this model. 
For simplicity we calculated our spectral model for the orbital
phase of maximum intensity ($\phase0.23-0.77$) to
match the phase coverage of the extracted spectrum.
We used three zones that cover the area between $\theta=0-20^\circ$, 
$20-50^\circ$ and $50-90^\circ$ on the surface of the companion.The size of the irradiated surface 
is determined by the relative radii of the stars and the semi-major axis of the system. The twilight zone
(transition region between the day and night sides) is less irradiated by the primary, here
we scaled the dilution factor by the relative visible area of the disk of the primary.
The night side does not contribute to the spectrum in this model.
Next, we calculated plane-parallel irradiated non-LTE model atmospheres 
with {\sc Tlusty} \citep{Hubeny1995} and synthetic spectra with 
{\sc Synspec} \citep{Lanz2007}, and reproduced the spectrum of the 
companion with a superposition of the irradiated concentric zones. 
The synthetic spectra of each zone were co-added with
relative weights: $0.08$, $0.37$ and $0.55$, that corresponds to 
the inclination of $89.21^\circ$ and considers
the phase-integrated contributions of the zones between $\phase0.23-0.77$.
All our models were calculated with opacity sampling with H, He and CNO 
composition and model atoms taken from the {\sc Tlusty} web 
page\footnote{\url{http://nova.astro.umd.edu/}}. 
We assumed that the irradiating flux is black-body like 
with $T_{\rm eff}=42\,000$\,K \citep{KleppyRauch2011}. 
The dilution factor describes the irradiation efficiency, 
that is, the fraction of the irradiating flux absorbed by the companion.
The dilution factor differs in our model from the geometric dilution factor 
\citep{mihalas78} as it depends on several other parameters as well.
It is affected by the system geometry, the irradiation angle as well as 
the spectral properties of the irradiating source. 
In our analysis, therefore, the dilution factor is a free parameter
that includes other poorly constrained or degenerate parameters.
The effective temperature of the irradiated layers in 
similar systems are reported to be in the range of $13\,000-15\,000$ K 
(e.g.~HW\,Vir, $T_{\rm sdB}=28\,000$ K, \citealt{Kiss2000}), 
which our models reproduce with a dilution factor of $0.021$. 
From the relative radii and distance of the stars 
the geometric dilution factor is 0.006 at the substellar point.
However, probably due to departures of the UV sdO spectrum 
from a black-body 
spectrum, the best model fit required a higher dilution 
factor of 0.06 in the $\theta=0-20^\circ$ zone. 
In the subsequent zones we used 0.049 and 0.021, 
which were scaled with $\cos{\theta}$.
Further details of the geometry can be found in \citet{Gunter2011}.

The model presented here is our first attempt to reproduce the spectrum of
the companion, therefore several simplifying 
approximations have been made.
The spectral model does not include a magnetic field, spots and clouds, and 
molecular opacities. 
We also neglected convection. While low mass stars are fully
convective, this approximation is valid for the bright side where irradiation
supresses convection \citep{podsi91}. 
Both stars are assumed to be spherical in a circular orbit. 
Our model does not include any horizontal energy transport 
that would distribute energy from the substellar point towards the night
side. 
Although such a meridional flow is plausible at strong irradiation 
and would
explain the homogeneity of the irradiated hemisphere \citep{Beer2002}, 
the current UVES observations are not suitable for such a study.

We fit the observed spectra in Figure\,\ref{fig:dMspectrum} with our 
iterative steepest-descent spectral fitting procedure 
{\sc XTgrid} \citep{Nemeth2012}. 
The structure of the irradiated atmosphere of the secondary star in \aador\
resembles to an
inverted B type star, and therefore it can be modelled with the same
techniques as hot stars. 
Based on the observed radius $R=0.11$ R$_\odot$, if the companion 
is a red dwarf its night side temperature is $\sim$2400 K, if it is 
below the hydrogen burning limit the temperature may be as low 
as $\sim$1700 K \citep{dieterich14}.
Therefore we fixed the effective temperature of the secondary 
at $T=2000$ K and the surface gravity had a starting value of 
$\log{g}=5.25$ cm\,s$^{-2}$, 
that is consistent with the photometric value \citep{Hilditch2003}.
The gravity, abundances and the dilution factor were
adjusted iteratively for the substellar point and the dilution factor was 
scaled geometrically in the subsequent zones. 
We assumed a homogeneous abundance distribution in the atmosphere and 
adjusted the abundances globally for the entire irradiated hemisphere. 
We calculated three model atmospheres to the different irradiation zones 
and their corresponding synthetic spectra in every iteration. 
The zonal spectra have been normalised, weighted and co-added and the final
spectrum shifted to the data as shown in Figure \ref{fig:TlustydMspectrum}. 
Our fit was based on the Balmer and metal emission lines, which explains 
the virtually negative flux in the figure.
Data points, such as the Balmer line self-absorption cores and far line 
wings, where the synthetic spectrum reaches negative values 
(below the continuum level) were taken out of the chi-square calculation by
the fit procedure. 
We applied a $20$ km s$^{-1}$ rotational broadening to 
approximate the smearing caused by the assumed 
tidally locked rotation of the 
companion and extraction of the spectrum in velocity space. 
The rotational broadening profile is symmetric only near the
secondary eclipse.
We found that, while a single irradiation zone with different dilution factors and 
system geometries can 
reproduce parts of the spectrum, only a multi-zone model could address 
the majority of the features simultaneously. 
We conclude, in agreement with \citet{Gunter2011}, that a single 
temperature model cannot reproduce the spectrum of the irradiated 
hemisphere in detail, even though Figure \ref{fig:Kcorr} suggests that 
the hot spot is largely uniform and covers the entire inner 
hemisphere of the secondary. 
The models are very insensitive to the effective temperature of the
companion, but the Balmer lines are sensitive to the surface gravity and the
dilution factor. 
This result is not surprising because the atmospheric temperature
structure is dominated by the irradiation that penetrates below the observed 
photosphere. 
The Balmer profiles are broad and shallow in the substellar
point and become narrower and stronger toward the terminator, therefore 
the line profiles depend also on the geometry and surface integration, which
makes a precise surface gravity determination complicated.  
We found surface gravities in the range of $\log{g} = 5.05-5.45$
cm\,s$^{-2}$ in the three zones.  The surface gravity of the companion can be precisely 
estimated in eclipsing binaries from the orbital 
parameters \citep{southworth04}. 
Using the results of \citet{Hilditch2003} and
the radial velocity amplitude 
derived in Section 4, we found $\log{g}=5.276\pm0.002$ cm
s$^{-2}$, although this method does not take irradiation effects into
account. 
The observed weak absorption in the wings 
of Balmer lines indicate that the surface gravity may depend on the
distance from the substellar point, which needs further investigations. 

The parameters of our best fit model are 
listed in Table \ref{tabl:2}.
Based on the measured CNO abundances we estimate the
metallicity of the
secondary to be $[M/H]=-0.7$ dex. 
This metallicity is surprisingly high and most likely not reflecting the
primordial metallicity of the companion. 
The C and O rich, N depleted abundance pattern resembles the sdO
primary \citep{Fleig2008}. 
Accretion from a radiatively driven optically thin wind of the sdO
\citep{krticka10} is a plausible reason
for the observed high metallicity and needs further investigations.
The consistent metal and Balmer line radial velocities exclude a circumstellar
origin of the metal lines. 

\begin{figure}
\centering
\includegraphics[angle=0, width=\hsize]{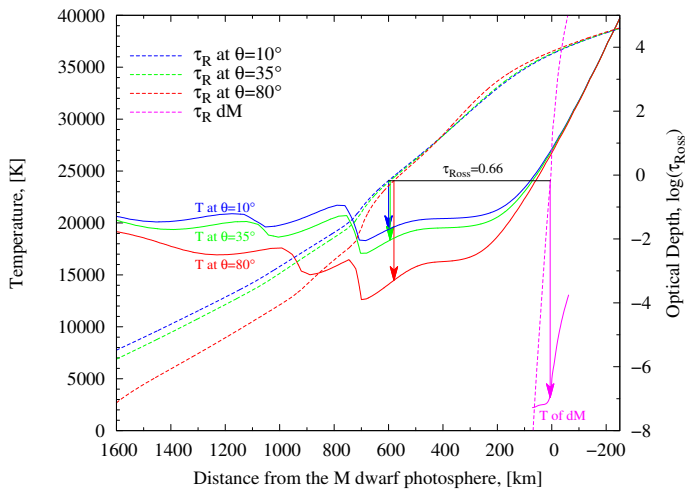}
\caption{
Temperature structure of the atmosphere of the irradiated dM secondary 
star in the \aador\ system. 
The three irradiated zones for 
$\theta=10^\circ$, $35^\circ$ and $80^\circ$ are shown together with the 
structure of the unperturbed dM on the night side. 
The irradiated zones show several temperature inversion regions 
and an extension of the atmosphere. 
The photospheres are progressively higher and hotter toward the substellar
point with respect 
to the dark side as indicated by the arrows at $\tau_{\rm Ross}=0.66$.  
}
\label{fig:TdM}
\end{figure}

\begin{table}\small
\begin{center}
\setstretch{1.3}
\begin{minipage}[h]{\linewidth}
\caption{Atmospheric parameters of the irradiated M dwarf in AA\,Dor.
Surface abundances are also listed with respect to the solar abundances
\citep{asplund09}.
\label{tabl:2}}
\begin{tabular}{lcccc}
\hline\hline
\multicolumn{2}{l}{}                              & Zone 1            & Zone 2                     & Zone 3              \\
\multicolumn{2}{l}{Parameter}                     & $\theta<20^\circ$ & $20^\circ<\theta<70^\circ$ & $70^\circ<\theta$   \\
\hline
\multicolumn{2}{l}{$T_{\rm eff}$ (K) day side} & 19200 & 18300 & 14300    \\
\multicolumn{2}{l}{Dilution factor}               & 0.060 & 0.049 & 0.021 \\
\hline
\multicolumn{2}{l}{$T_{\rm eff}$ (K) night side} & \multicolumn{3}{c}{$2000\pm500$ } \\
\multicolumn{2}{l}{$\log g$ (cm s$^{-2}$)}              &
\multicolumn{3}{c}{$5.25\pm0.25$ } \\
\multicolumn{2}{l}{F$_{\rm dM}$/F$_{\rm sdO}$ (at 5000 \AA,
\phase0.47)} & \multicolumn{3}{c}{ $0.07\pm0.02$  }  \\
\hline\hline
\multicolumn{1}{l}{Abundances}      &$\log(n{\rm X}/n{\rm H})$ &$+$ & $-$ & $\times$ Solar\\
\hline
\multicolumn{1}{c}{He}              &-1.69   &0.88 & 0.38&0.24\\
\multicolumn{1}{c}{C}               &-4.30   &0.02 & 0.11&0.19\\
\multicolumn{1}{c}{N}               &-4.89   &0.06 & 0.04&0.19\\
\multicolumn{1}{c}{O}               &-3.94   &0.01 & 0.03&0.23\\
\hline\hline
\end{tabular}
\label{Tab_xtgrid}
\end{minipage}
\end{center}
\end{table}

{\sc Tlusty} provides the atmospheric temperature, density and pressure
structure that allowed us to estimate the extension of the atmosphere. 
Figure\,\ref{fig:TdM} shows the vertical temperature structure of the three
irradiated zones.
We found inversion layers where temperature changes rapidly with
optical depth. 
By stacking plane-parallel models with {\sc XTgrid} we could account 
for sphericity effects. 
The optical depth scale was 
integrated inward and synthetic spectra were calculated at subsequently 
higher layers towards the limb. 
The differences among the zones in photospheric depths are relatively small. 
We found that the pressure broadened Balmer emission line wings 
form deeper than the line core and most of the metal lines. 
The steep temperature gradients in the  
inversion layers and line forming region provide a 
spectral diversity to the zones. 
With {\sc CoolTlusty}, an optimised version of the code for brown dwarf and planetary
atmospheres \citep{Hubeny2003}, we calculated a dM model for the
night side of
the companion. 
The pressure scale height was found to be $8$ km for the night side
and $130$ km in the substellar zone, suggesting an extended atmosphere.
Interestingly, these scale heights are comparable to that of the 
Earth and the Sun. 
By matching the photospheric pressure of the M dwarf on the dark side 
with the pressure in the irradiated zones we shifted the zones to the 
same base. 
This allowed us to measure the thickness of the atmosphere in the zones. 
We found that the irradiated zones have a larger radius by about 600\,km, 
which is about 0.8 \% of the radius of the secondary. 
Similar expansions have been observed in irradiated extrasolar giant
planets \citep{Burrows2007}.

\begin{figure*}[ht!]
\centering
\includegraphics[angle=0,width=\hsize]{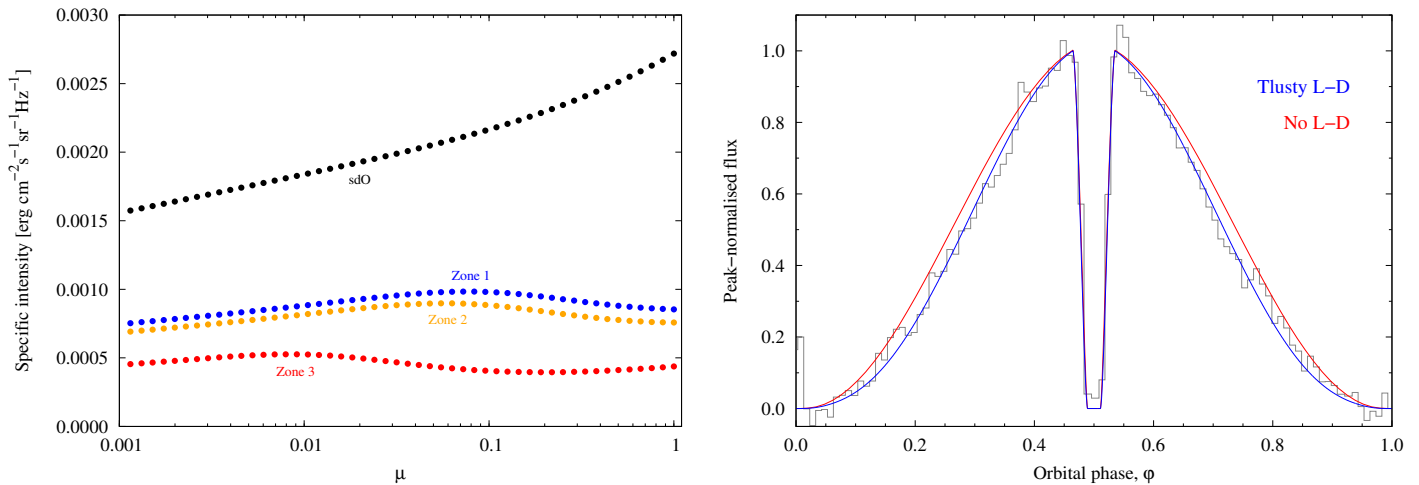}
\caption{ {\it Left:} Specific intensities for the sdO star and for 
the three irradiated zones on the M dwarf. 
Note, we use a logarithmic scale on the abscissa to better represent the center to
edge variations of the specific intensities, $\mu$ is the cosine of the polar 
angle (between the line of sight and the surface normal at a given point 
on the companion) that takes the value 1 in the disk center and 0 on the edge.
There is only a little difference between the 
substellar zone (No. 1) and the intermediate zone (No. 2), but the
terminator zone (No. 3) is substantially fainter. 
Each zone shows a moderate brightening towards the limb. 
{\it Right:} Effects of limb-darkening on the phase profile. The three zone 
{\sc Tlusty} model provides only a marginally better fit than a 
flat surface intensity distribution of the companion.
}
\label{fig:LD}
\end{figure*}

From the spectral models the limb-darkening of the irradiated 
hemisphere can be evaluated numerically. 
The left panel of Figure \ref{fig:LD} shows the 
limb-darkening of the three irradiated zones
and of the sdO star for comparison. 
The actual surface flux distribution is not only function of the angle to the
surface normal, but also depends on the orbital phase and distance from the
substellar point.
Therefore, we developed a 
model that allows us to use the specific intensities and 
surface integration in the phase profile calculation. 
With this model the total surface intensity of the visible part of the 
irradiated hemisphere can be calculated for an arbitrary orbital phase and 
the light curve of the secondary can be obtained. 
We used again the published photometric radii, inclination and semi-major axis 
of the system \citep{Hilditch2003}. 
The right panel of Figure \ref{fig:LD} shows that the model 
reproduced the observed phase 
curve quite well. 
The model also reproduced the size of the spot, that can be 
inferred from the phase coverage of the reflection effect, 
in good agreement with the
analysis in Section \ref{sect:K2modeling}. 
The phase curve shows some extra flux between $\phase0.75-0.95$.
We interpret this excess flux as the signature of an asymmetry in the hot
spot due to a non-synchronous rotation of the companion with the orbit. 
The Roche potential of the companion does not suggest a notable deformation,
but a small atmospheric expansion near the substellar point is likely 
due to irradiation.
However, its effect on the phase curve must be negligible as an ellipsoidal-like
variation would give a maximal 
contribution at $\phase0.25$, which is not obvious in Figure \ref{fig:LD}.

To overcome the limitations of our current models we will focus on three
major improvements in the future: 
(1) replacing the black-body irradiation with the spectrum of
the sdO star; (2) implementation of a more adjustable surface intensity distribution
that will allow us to investigate the asymmetries of the hot spot and (3)
improving the consistency between
the spectral and light curve models with the help of optical-infrared
photometry. 
A better consistency may also require more irradiation zones and opacity sources. 
Such a model will allow us to interpret high precision light 
curves and infer the conditions in the atmospheres of irradiated 
companions in faint reflection effect
binaries where direct spectroscopy is not yet feasible.  
We will present our progress in a forthcoming publication.

\section{Discussion and summary}\label{sect:discussion}

We have reanalysed the UVES dataset of the \aador\ binary system, with particular attention to extracting orbital phase information from the emission lines of the irradiated secondary. We integrated the signal from the Balmer lines and 22 narrow emission lines, and were able to firmly establish that the emission region covers no less than 95\% of the irradiated hemisphere. With this information we were able to accurately translate the observed center-of-light velocity, \kirr, into a precise center-of-mass velocity, \kdm. With this final piece of the puzzle in place, we were able to finally establish the masses of the two components with very high precision. The new numbers are fully compatible with the results of \citet{Hoyer2015}, but with error bars that are 5-10 times lower. 

With radial velocities of the primary, \ksd\,=\,--39.63\,$\pm$\,0.21\,km/s and the secondary, \kdm\,=\,231.3 $\pm$ 0.7\,km/s we get the mass for the primary,  \msd\,=\,0.46 $\pm$ 0.01\,\msol, and the secondary, \mdm\,=\,0.079 $\pm$ 0.002\,\msol.
These masses are in full agreement with \citet{KleppyRauch2011} and \citet{Hoyer2015}, confirming the hot subdwarf primary star and tightly confining the secondary to the hydrogen burning mass limit, i.e. either a brown dwarf or a late dM star \citep{Baraffe2015}.

To further study the nature of the secondary star in \aador, we need to obtain high-S/N near-infrared (NIR) spectra from both the cold and hot hemisphere. The recent attempt to obtain such data by  \citet{Hoyer2015} was not successful as poor observing conditions prevented them from detecting any signal from the secondary in the NIR part of the XSHOOTER time-resolved data set. We strongly encourage repeating this attempt in excellent observing conditions. 

Combining the 22 metal lines of the secondary to study the shape of the emission velocity profiles we detected two distinct asymmetry features. The velocity profiles appear skewed at the quadratures which, due to the pure geometrical effect, produces a sort of a self-eclipse. The asymmetry in the velocity profiles emerging at the secondary eclipse is caused by the rotation of the secondary and hence we termed it as the RM effect of the hot spot.  

Furthermore, we have modelled the atmosphere of the secondary of \aador\ including the irradiation effects that matched the observed spectrum of the companion very well. Comparing the synthetic and observed spectra we confirmed the majority of the emission lines as listed in Table~\ref{tbl:emlines}. While the hot spot covers almost the entire inner hemisphere, a single temperature model fails to reproduce the observed irradiated spectrum indicating that, while there must be a very effective process distributing the energy over the irradiated hemisphere, the irradiated regions contribute differently to the spectrum of the companion. The irradiated zones show several temperature inversion regions 
and an extension of the atmosphere.

We developed a model that allows us to use a surface integration of the specific intensities in the phase profile calculation. With this model the total surface intensity of the visible part of the hot 
spot can be calculated for an arbitrary orbital phase and the light curve of the secondary can be obtained.

The observations of irradiated planets are indicative of the heat transport from the day-side to the night-side \citep{Knutson2007} however, to understand the circulation patterns of hot Jupiters it is crucial to analyse the temperature inversion in the upper atmospheres. While we are witnessing the era of hot Jupiters and circumbinary planets around evolved binaries,  \aador\  stands as a bona fide system to understand the radiation transport in the extreme conditions of irradiated (sub)stellar objects.

\begin{acknowledgements}
We thank Cedric Ledoux for helpful discussions on UVES data reduction.
We thank Tom Marsh for the use of his spectrum analysis program \texttt{Molly}.
This research was based on observations collected at the European Southern Observatory, (Chile) Program ID: 066.D-1800. This research has made use of SIMBAD database, operated at the CDS (Strasbourg, France) and NASA  Astrophysics Data System Bibliographic Services. 
\end{acknowledgements}

\bibliographystyle{aa} 
\bibliography{references} 

\Online
\begin{appendix}
\section{Emission lines from the secondary star}

Table \ref{tbl:emlines} gives a list of all emission lines from the 
secondary star of the \aador\ system we detected in the UVES spectra. 
Note that only the first 22 lines were used in the analysis. 

\begin{table}[ht]
\caption[]{List of emission features from the secondary.}
\label{tbl:emlines}
\centering
\begin{tabular}{llccc} \hline\hline
Ion & Wavelength  [\AA]& Peak & EW [m\AA]&\\ \hline
\ion{C}{ii} & 3920.68 
& 0.01051 & -3.5  &\\ 
\ion{O}{ii} & 3954.37 
&  0.00891 & -2.3 &\\ 
\ion{O}{ii} & 4069.75 
& 0.01272 & -4.2  &\\ 
\ion{O}{ii} & 4072.16 
& 0.01288 & -2.2  &\\ 
\ion{O}{ii} & 4075.87 
& 0.01579 & -4.6 & \\ 
\ion{O}{ii} & 4089.29 
& 0.00893 & -2.4  &\\ 
\ion{O}{ii} & 4119.22 
& 0.01426 & -4.5  &\\ 
\ion{O}{ii} & 4153.30 
& 0.00746 & -2.3 &\\ 
\ion{O}{ii} & 4189.79 
& 0.00917 & -3.0  &\\ 
\ion{C}{ii} & 4267.10 
& 0.02787 & -13.4 &\\ 
\ion{O}{ii} & 4414.91 
& 0.02147 & -7.1 & \\ 
\ion{O}{ii} & 4416.97 
& 0.01617& -5.6  &\\ 
\ion{O}{ii} & 4590.97 
& 0.01206 & -4.5  &\\ 
\ion{O}{ii} & 4596.17 
& 0.00883 & -2.7  &\\ 
\ion{O}{ii} & 4641.81 
& 0.01402 & -4.4   &\\ 
\ion{C}{iii} & 4647.40 
& 0.01114 & -2.9  & \\ 
\ion{O}{ii} & 4649.14 
& 0.01387 & -4.0 &  \\
\ion{O}{ii} & 4661.64 
& 0.00951& -2.8 & \\ 
\ion{O}{ii} & 4676.23 
& 0.00984 & -2.6  & \\ 
\ion{O}{ii} & 4699.21 
& 0.00951 & -3.7 &  \\ 
\ion{O}{ii} & 4705.35 
& 0.01355 & -4.9 & \\ 
\ion{O}{ii} & 4943.06 
& 0.01050 & -2.6 & \\ 

\hline
\multicolumn{5}{l}{Lines detected, but not used in the analysis}\\
\hline
\ion{He}{i} &\multicolumn{4}{l}{(3819.61+3819.76), (3867.48+3867.63), 3871.82,} \\
            &\multicolumn{4}{l}{3935.91,3964.73,(4026.19+4026.36),} \\
            &\multicolumn{4}{l}{4120.81,4143.76,4168.97,4387.93,4471.69,} \\
            &\multicolumn{4}{l}{(4713.14+4713.37), 4921.929}\\
\ion{Si}{iii}&\multicolumn{4}{l}{3806.56,4552.65,4567.87} \\
\ion{Ca}{ii} &\multicolumn{4}{l}{3933.66} \\
\ion{Mg}{ii}&\multicolumn{4}{l}{4481.13} \\
\ion{N}{ii} &\multicolumn{4}{l}{3994.99, 3998.69, 4035.09, 4432.74, 4447.03, 4552.54} \\
\ion{C}{ii} &\multicolumn{4}{l}{(3876.05+3876.19+3876.41), 3918.98} \\
                &\multicolumn{4}{l}{(4411.20+4411.52)} \\
\ion{C}{iii}&\multicolumn{4}{l}{4647.40, 4650.16} \\
\ion{O}{ii}&\multicolumn{4}{l}{3882.20, 3919.29, 3945.05, 3973.26, 3982.72, 4035.09, }\\
           &\multicolumn{4}{l}{4060.58, 4078.86, 4085.12, 4087.16, } \\
           &\multicolumn{4}{l}{4092.94, 4097.26, 4119.22, 4132.81, 4185.46, 4192.50, } \\
           &\multicolumn{4}{l}{(4253.74+4253.98), 4275.52, 4303.82, } \\
           &\multicolumn{4}{l}{4317.14, 4319.63, 4342.00, 4342.83, 4343.36, 4344.42, } \\
           &\multicolumn{4}{l}{4345.56, 4347.42, 4349.43, 4351.27, 4353.60, 4366.90, } \\
           &\multicolumn{4}{l}{4369.28, 4378.41, 4395.95, 4448.21, } \\
           &\multicolumn{4}{l}{4452.38, 4466.32, (4609.42+4610.14), } \\
           &\multicolumn{4}{l}{4638.85, 4649.14, 4650.84, } \\
           &\multicolumn{4}{l}{4696.36, (4701.23+4701.76), 4703.18, } \\
           &\multicolumn{4}{l}{4710.04, 4906.88, 4924.60, 4941.12} \\ \hline
\end{tabular}
\end{table}

\end{appendix}

\end{document}